\documentclass{PoS}

\usepackage{slashed}
\usepackage{color, verbatim}
\usepackage{latexsym}
\usepackage{epsfig}
\usepackage{amsmath,accents}

\usepackage{amssymb}
\usepackage{graphicx}
\usepackage{bm}

\title{\centering LISA as a probe for particle physics:\\electroweak scale tests in synergy with ground-based experiments}

\ShortTitle{LISA as a probe for particle physics}

\author{D.~G.~Figueroa$^{1}$, E.~Meg\'ias$^{2, 3}$,  
        \speaker{G.~Nardini}${\,}^{4,5}$,  M.~Pieroni$^{6}$, M.~Quiros$^{7}$, A.~Ricciardone$^{8}$, G.~Tasinato$^{9}$\\
        \llap{$1$} Institute of Physics, Laboratory of Particle Physics and Cosmology, \'Ecole Polytechnique F\'ed\'erale de Lausanne, CH-1015 Lausanne, Switzerland \\
       \llap{$^2$}
       Departamento de F\'{\i}sica At\'omica, Molecular y Nuclear and  Instituto Carlos I de F\'{\i}sica Te\'orica y Computacional, Universidad de Granada, E-18071 Granada, Spain \\
        \llap{$^3$} Departamento de F\'{\i}sica Te\'orica, Universidad del Pa\'{\i}s Vasco UPV/EHU, Apartado 644,  E-48080 Bilbao, Spain \\
        \llap{$^4$}
        AEC, ITP,
        University of Bern, CH-3012 Bern, Switzerland
        \\
        \llap{$^5$} 
        Department of Physics and Mathematics, University of Stavanger, N-4036 Stavanger, Norway
        \\
        \llap{$^6$} Instituto de F\'{\i}sica Te\'orica UAM/CSIC, Universidad Aut\'onoma de Madrid Cantoblanco, Madrid 28049, Spain
        \\
        \llap{$^7$}  Institut de F\'{\i}sica d'Altes Energies \& BIST, Universidad Autonoma de Barcelona, E-08193 Bellaterra (Barcelona), Spain
       \\
	\llap{$^8$} Department of Mathematics and Physics, University of Stavanger, 4036, Stavanger, Norway \\
        \llap{$^9$} College of Science, Swansea University, Swansea, SA2 8PP, UK \\
       \\
        E-mail: 
         \email{daniel.figueroa@cern.ch}, 
        \email{emegias@ugr.es}, 
        \email{nardini@itp.unibe.ch},
        \email{mauro.pieroni@uam.es},
        \email{quiros@ifae.es},
	\email{angelo.ricciardone@uis.no},
        \email{g.tasinato@swansea.ac.uk}
        }

      \abstract{We  forecast the  prospective of detection for a stochastic gravitational wave background sourced by cosmological first-order phase transitions. We focus on first-order phase transitions with negligible plasma effects, and consider the experimental infrastructures built by the end of the LISA mission. We make manifest the  synergy among LISA, pulsar time array experiments, and ground-based interferometers. For phase transitions above the TeV scale or below the electroweak scale, LISA can detect the corresponding gravitational wave signal together with Einstein Telescope, SKA or even aLIGO-aVirgo-KAGRA. For phase transitions at the electroweak scale, instead, LISA can be the only experiment observing the gravitational wave signal. In case of detection, by using  a parameter reconstruction method that we anticipate in this work, we show that   LISA on its own has the potential to determine when the phase transition occurs and, consequently, the energy scale above which the standard model of particle physics needs to be modified. The result may likely guide the collider community in the post-LHC era.}

\FullConference{GRAvitational-waves Science\&technology Symposium - GRASS2018\\
		1-2 March 2018\\
		Palazzo Moroni, Padova (Italy)}

\begin{document}

%
\section{Introduction}

Several phenomena in the early universe generate a Stochastic Gravitational Wave Background (SGWB) that can be detected by current
   or forthcoming Gravitational Wave (GW) experiments~\cite{Caprini:2015zlo, Bartolo:2016ami, Caprini:2018mtu}. The observation of a GW signal associated to any of these sources would be a milestone in the understanding of nature. It would be the first \emph{direct} measurement of the properties of the universe before the cosmic microwave background fingerprint, and  would put on firm observational ground an epoch prior to Big Bang Nucleosynthesis (BBN). 

The SGWB  associated to a cosmological First-Order Phase Transition (FOPT) is certainly one of the most investigated. The Standard Model of particle physics (SM) predicts that no cosmological FOPT has ever occurred~\cite{Kajantie:1996mn}. Nevertheless, many theoretical and empirical observations (e.g.~the hierarchy problem, the dark matter, and the baryon-asymmetry of the universe) suggest that the SM cannot be the ultimate theory, and many completions of the SM have been proposed to solve some of its theoretical  issues. Remarkably, several of them predict a cosmological FOPT SGWB in the LISA frequency band~\cite{Caprini:2015zlo}.

The spectrum of GW energy density per logarithmic frequency interval of a SGWB  is defined as
$\Omega_{\rm GW}(f)\equiv \rho_c^{-1} {d\rho_{\rm GW}}/{d(\log f)}\,$,
with $d \rho_{\rm GW}$ being the GW energy density in the interval $f$ and $f+df$, and  
$\rho_c = 8\times 10^{-47} h^2\,$GeV$^4$ the critical density normalized by the Hubble scale factor $h\simeq 0.67$. For a cosmological FOPT, this spectrum can be expressed in terms of a few effective parameters that describe the particle content of the plasma and its interactions with the fields triggering the phase transition. For simplicity hereafter we focus on setups where such interactions are negligible and the FOPT has substantial supercooling~\footnote{The analysis we are presenting could be straightforwardly repeated for scenarios with small supercooling or plasma contributions. The only complication is the larger dimensionality of the effective-parameter spaces to study.}. In this case the SGWB spectrum  can be written as~\cite{Caprini:2015zlo, Huber:2008hg, Cutting:2018tjt} 
\begin{eqnarray}
h^2\Omega_{\rm GW}(f) &\simeq&  
h^2\overline \Omega_{\rm GW} ~\frac{3.8(f/ f_p)^{2.8}}{1+2.8 (f/ f_p)^{3.8}}  \,,
\label{eq:OmGW}
\end{eqnarray}
where the (normalized) amplitude and frequency of the peak of the signal are respectively
\begin{equation}
h^2\overline \Omega_{\rm GW}\simeq 
\frac{8}{10^{5}} 
\left( \frac{H_\star}{\beta}  \right)^2 \frac{\xi(v_w)}{\sqrt[3]{g_\star}} \,, 
\qquad
f_p\simeq \frac{7.7}{10^{5}} ~\tilde{\xi}(v_w) \left(\frac{\beta}{H_\star}\right)
\frac{T_\star\sqrt[6]{g_\star(T_\star)}}{100\, \textrm{GeV}}~ \textrm{Hz} \,,
\label{eq:fp}
\end{equation}
with
\begin{equation}
\xi(v_w)=\frac{0.11 v_w^3}{0.42+v_w^2} ~, \qquad  \tilde \xi(v_w) = \frac{0.62}{1.8-0.1+v_w^2}~.
\end{equation}
The frequency shape of the spectrum thus depends on various parameters: the velocity at which the bubble collide ($v_w$); the temperature at which the FOPT ends ($T_\star$); the inverse of the FOPT duration normalized by the Hubble factor at the completion of the transition ($\beta/H_\star$); and the number of relativistic species in the thermal bath after the transition ($g_\star(T_\star)$). A FOPT occurring during the electroweak symmetry breaking typically exhibits $T_\star$ of the order of 100\,GeV, while the values of $\beta/H_\star$ and $v_w$ are very model dependent. Most of the SM extensions provide  $g_\star(T_\star\simeq 100\,\textrm{GeV})\approx 106$. Nevertheless, there are no model-independent experimental constraint preventing FOPT at temperatures far away from the electroweak scale, so that the frequency of the peak, $f_p$, can span many orders of magnitude, as shown in Fig.~\ref{fig:sens} (left panel). Fig.~\ref{fig:sens} also shows
$h^2\Omega_{\rm GW}(f)$ for some illustrative FOPT scenarios (dotted curves in the right panel).

LISA is particularly sensitive, but not limited, to FOPT at the electroweak scale. Due to the frequency broadband of $\Omega_{\rm GW}(f)$, LISA can detect FOPTs with $10^{-3}\, \textrm{GeV} \lesssim T_* \lesssim 10^7$ GeV and $\beta/H_\star$ up to $10^3$ when bubbles are ultrarelativistic, $v_w \simeq 1$~\cite{Caprini:2015zlo}. For the same reason, a FOPT signal detected at LISA can in principle be observed at the most diverse frequencies. On the other hand, depending on the FOPT parameter region, other experiments can have better chances to detect the signal. In view of this feature we now estimate the parameter reach of the  current and forthcoming GW network (Section~\ref{sec:synergy}), sketch how well LISA can reconstruct the FOPT signal (Section~\ref{sec:recontr}) and comment on the implications that such a reconstruction has on the whole GW detector network (Section~\ref{sec:concl}).  

%
\section{Synergy between different GW observatories}
\label{sec:synergy}

\begin{figure}[t]
\centering
\includegraphics[width=6.6cm]{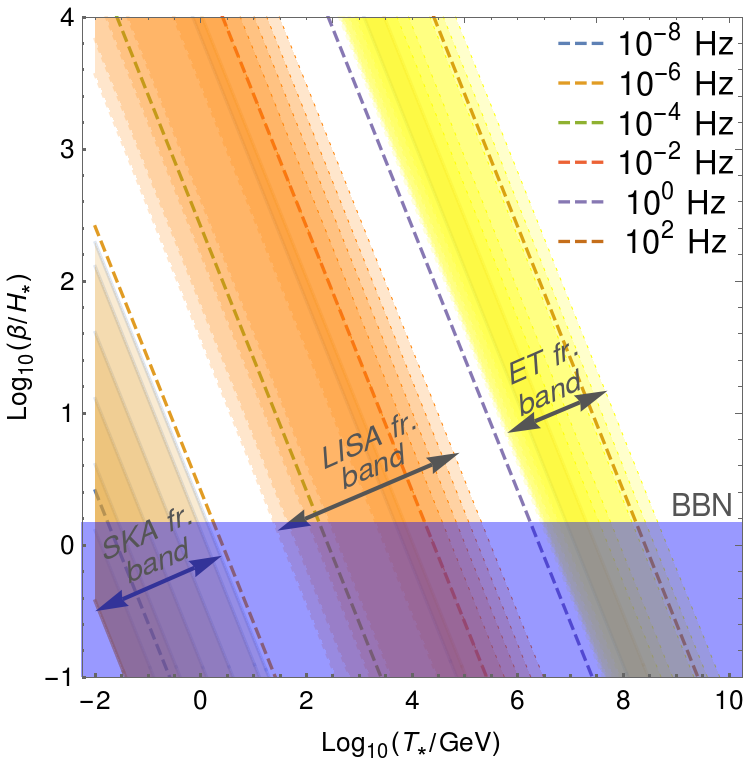} \hspace{0.5cm}
\includegraphics[width=7cm]{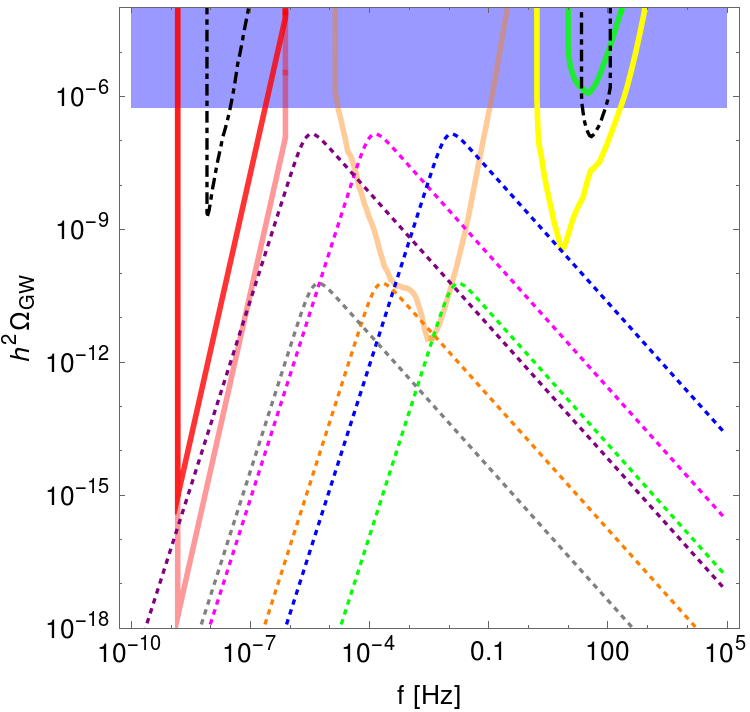}
\caption{\it
Left panel: The value of the frequency peak $f_p$ in the parameter space $T_\star$--$\beta/H_\star$ for bubble velocities $v_w=0.99$. On the background (colored bands) the frequencies the forthcoming GW detectors are sensitive to; the stronger the color, the better the sensitivity. The blue area is excluded by the BBN bound.
Right panel: Sensitivity curves of the current and forthcoming GW experiments and the SGWB signals (dotted curves) sourced by the  FOPT benchmark scenarios with $g_\star = 106$, $\beta/H_\star=3$, and $T_\star/\,$GeV and $v_w$ being respectively 3 and 0.99 (dotted purple line), 3 and 0.05 (dotted gray line), 120 and 0.99 (dotted magenta line), 120 and 0.05 (dotted orange line), 10$^4$ and 0.99 (dotted blue line), and 10$^4$ and 0.05 (dotted green line). 
The dotted-dashed lines correspond to the power-law sensitivity curves of  PPTA \& EPTA \& NANOGRAV (at frequencies $f\sim$~nHz) and aLIGO O1 (at frequencies $f\sim$~100\,Hz); the solid lines correspond to the sensitivity curves $\Omega_{\rm sens}(f)$ of SKA observing 100 milli-second pulsars (dark red), SKA observing 2000 milli-second pulsars (light red), LISA (orange), ET (yellow), and aLIGO-aVirgo-KAGRA network at its final design (green). The BBN bound rules out the FOPT SGWB signals touching the blue area.}
\label{fig:sens}
\end{figure} 

\begin{figure}[t]
\centering
\includegraphics[width=7cm]{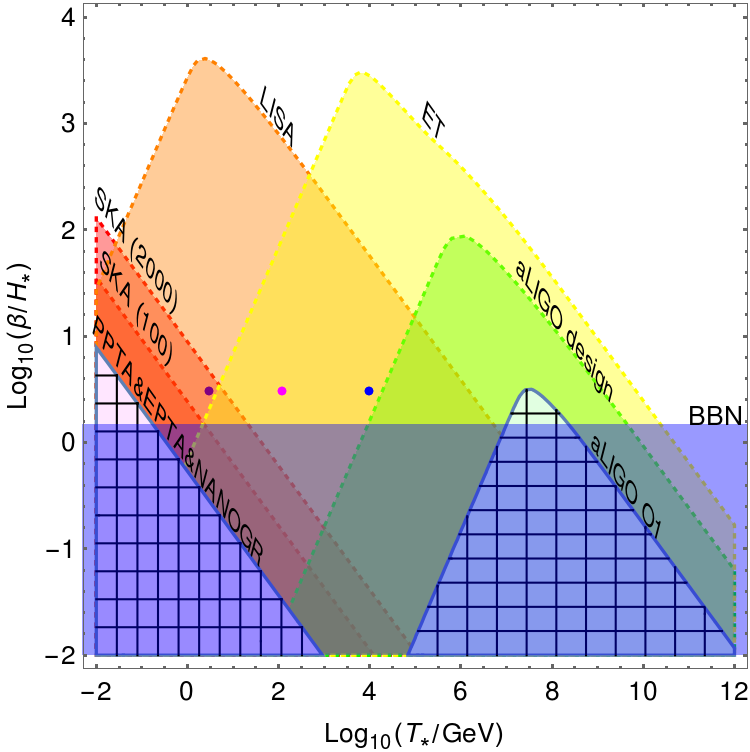} \hspace{0.5cm}
\includegraphics[width=7cm]{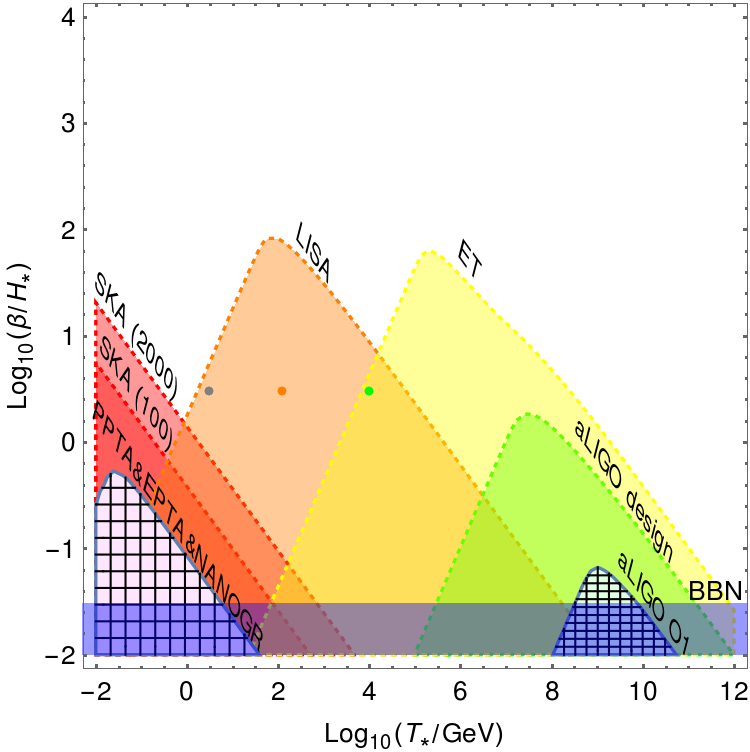} 
\caption{\it 
The parameter reach of the future GW experiment network for $v_w=0.99$ (left panel) and $v_w=0.05$ (right panel). In each area the corresponding experiment (see labels) detect the FOPT with SNR $> 10$. The BBN bound rules out the blue region. The bullet points correspond to the benchmark SGWB signals appearing with the same color in the right panel of Fig.~\ref{fig:sens}.
}
\label{fig:reach}
\end{figure}

Pulsar time array experiments such as PPTA, EPTA and NANOGRAV are currently testing the frequency band $\mathcal{O}(1)$--$\mathcal{O}(10^3)$ nHz, while present ground-based interferometers are probing the $\mathcal{O}(10)$--$\mathcal{O}(10^4)$ Hz range. The lack of detection in the collected data~\cite{Shannon:2015ect, Lentati:2015qwp, TheLIGOScientific:2016dpb, Arzoumanian:2018saf} rules out any SGWB with a power-law power spectrum (i.e.~$\Omega_{\rm GW}(f)= \Omega_0 f^n$ with $\Omega_0$ and $n$ constants) that intersects the \textit{power-law sensitivity curves} displayed in Fig.~\ref{fig:sens} (right panel, dotted-dashed curves). The lack of detection can be converted into an exclusion bound on $h^2\Omega_{\rm GW}(f)$ and the corresponding FOPT effective parameters. This bound rules out the etched parameter areas in Fig.~\ref{fig:reach}.

In the future, the two aforementioned frequency ranges will be partially extended with SKA and Einstein Telescope (ET), and most of the intermediate frequencies will be covered by LISA. The right panel of Fig.~\ref{fig:sens} displays the (non power-law) sensitivity curves of SKA (dark red and light red), LISA~\cite{Audley:2017drz} (orange), aLIGO in its final design~\cite{Aasi:2013wya} (green), and ET (yellow)~\cite{Sathyaprakash:2012jk}. For the two SKA curves we follow Ref.~\cite{Moore:2014lga} and assume observation of 100 (dark red) or 2000 (light red) milli-second pulsars~\cite{Megias:2018sxv}. 
For the forecast of the FOPT parameter reach we consider GW data that will be presumably collected by the (nominal) end of the LISA mission, in the late 2030s~\footnote{The LISA mission nominally ends around six years after its launch. Nevertheless, after this period, LISA should still be able to take data.}. 
The findings are shown in Fig.~\ref{fig:reach}. Different colored areas correspond to different detectors. Inside each area the associated experiment detects the FOPT SGWB with a Signal to Noise Ratio (SNR) larger than 10, with SNR given by
\begin{equation}
{\rm SNR_i} = \sqrt{(3.16\times 10^7 s)\frac{\mathcal T_i}{\textrm{1 year}}  \int_0^\infty df \frac{\Omega_{\rm GW}^2(f)}{\Omega_{\rm sens,i}^2(f)}     }~,
\end{equation}
with $\Omega_{\rm sens,i}(f)$ being the sensitivity curve of the $i$ experiment
having stored $\mathcal T_i$ years of data. Due to the uncertainties on the time schedules and duty cycles of all these detectors, it seems sensible to take $\mathcal T_i =$ 3, 7, 8, and 20 years for $i=$``LISA", ``ET" , ``aLIGO design", and ``SKA"  respectively, but of course the findings are indicative~\footnote{A similar analysis for Cosmic Explorer~\cite{Evans:2016mbw} and LISA can be found in Ref.~\cite{Axen:2018zvb}.}.

The BBN constraint, which limits the amount of radiation during the formation of the primordial elements, imposes an upper bound on the SGWB~\cite{Caprini:2018mtu} and, in turn, on $\overline{\Omega}_{\rm GW}$~\cite{Megias:2018sxv}. Due to this constraint, in the right panel of Fig.~\ref{fig:sens}, the FOPT SGWB power spectra entering the blue region are ruled out. Equivalently, the blue parameter areas in Fig.~\ref{fig:sens} (right panel) and Fig.~\ref{fig:reach} are experimentally excluded.  

The outcome of the forecast is remarkable: in around twenty years, if $\beta/H_\star \lesssim 10^4$, the full network of GW observatories will be able to test a substantial part of the FOPT effective parameter space. Moreover, in a wide fraction of this parameter space, at least two experiments will be able to detect independently the signal with SNR $>10$.

%
\section{Parameter reconstruction at LISA}
\label{sec:recontr}

\begin{figure}[t]
\centering
\hspace{-.2cm}
\includegraphics[width=7.5cm]{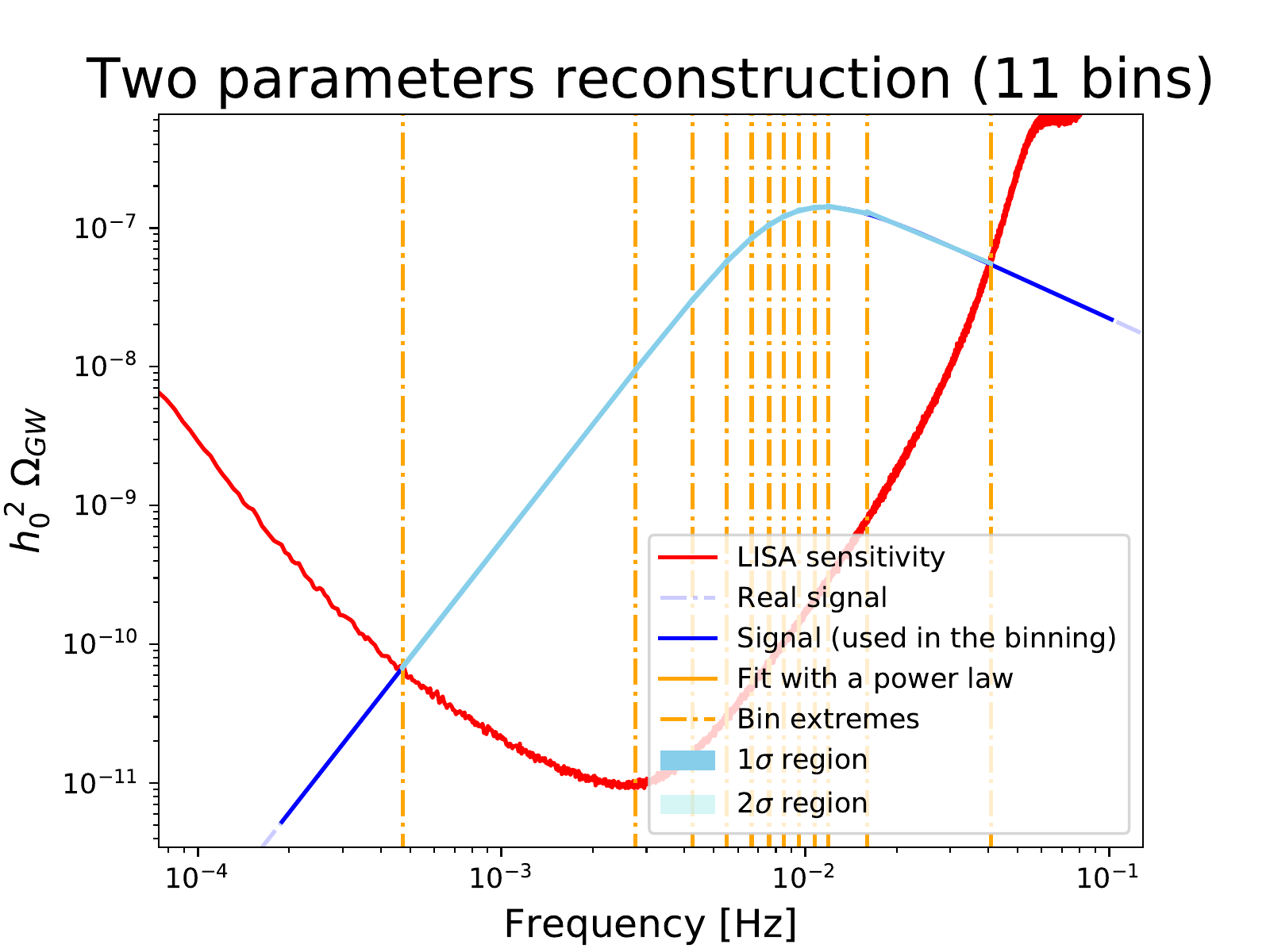}
%
\hspace{-.0cm}
\includegraphics[width=7.5cm]{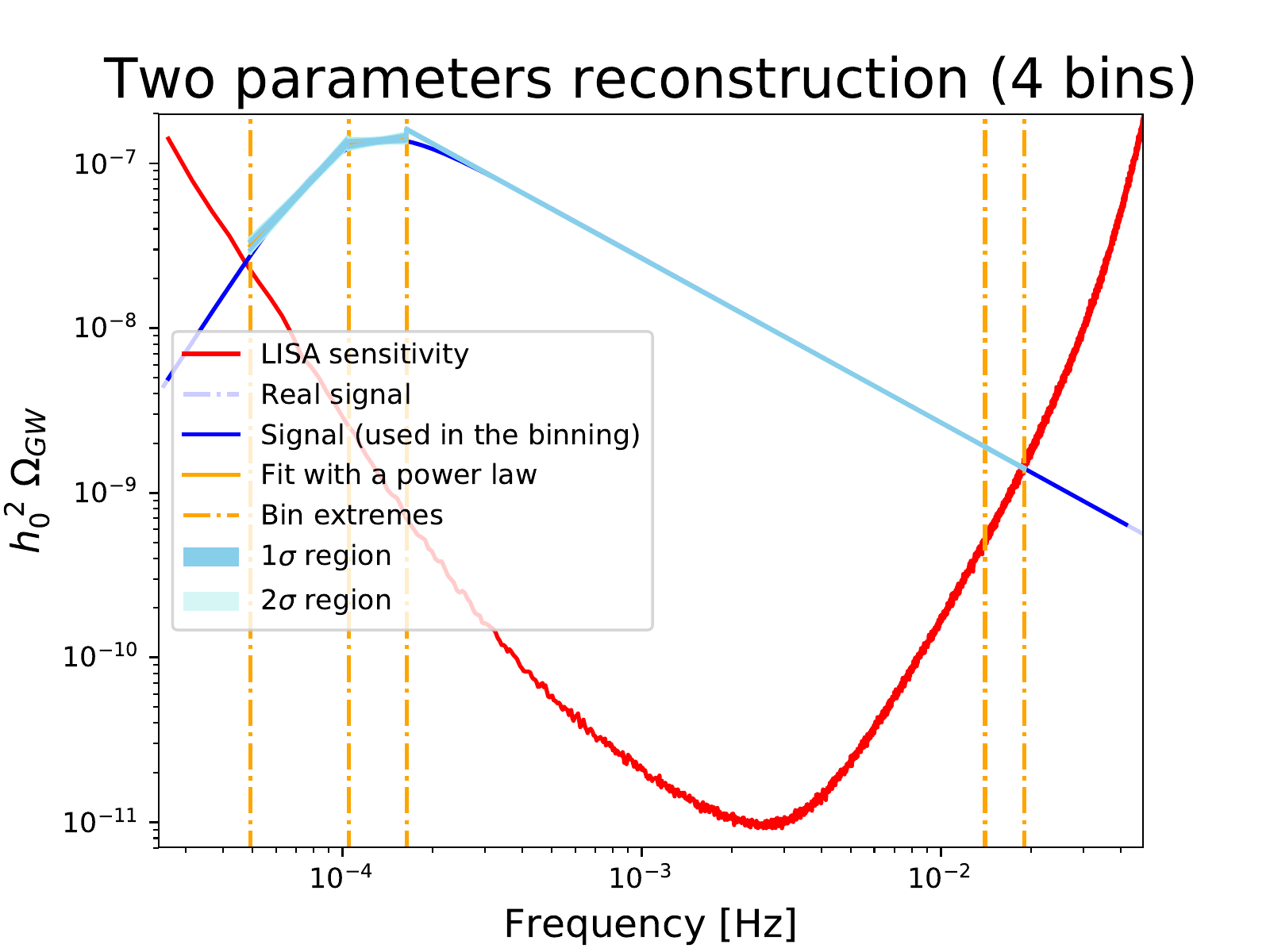} 
\hspace{-5mm}
\caption{\it
Reconstruction of the FOPT SGWB signals (blue lines) displayed in the left panel of Fig.~\ref{fig:sens} by means of the ``multi-bin" procedure assuming a good knowledge on the LISA instrumental noise (red curves). The power-law fit is performed inside several intervals indicated by vertical yellow dashed lines.  The 1-$\sigma$ uncertainties are represented by the light blue band. The signal reconstruct is performed only where the error bands are displayed.}
\label{fig:recons}
\end{figure}

As previously stated, several phenomena may have generated a SGWB. Therefore, given a SGWB detection, insights on the source are possible only by means of an accurate characterization of the signal. Obviously, in the case of the cosmological FOPT, the reconstruction of the frequency shape $\Omega_{\rm GW}(f)$ would be a key information. Here we anticipate how LISA data can be analyzed  to infer the frequency shape of a SGWB with high SNR~\cite{CosWG1}. For the present purposes, we apply this method to the FOPT benchmark signals with $v_w=0.99$ displayed in the right panel of Fig.~\ref{fig:sens} (magenta and blue curves).  

Those benchmark signals are not well described by a single power-law in the entire LISA frequency band.
Ref.~\cite{Adams:2013qma} investigates a parameter reconstruction for a SGWB whose frequency dependence is a power law. In particular, Ref.~\cite{Adams:2013qma} shows that  one can reconstruct the amplitude and slope  of the signal, if its SNR is of order 10 or larger. The SNR of our benchmark signals is several orders of magnitude larger than this value. Hence, we can split the total LISA frequency band in small subintervals -- the bins -- each characterized by sufficient SNR,  and accurately reconstruct  within each bin
the signal frequency shape in terms of a power-law~\cite{CosWG1}. Optimization and iteration of this procedure leads to the reconstruction of a SGWB in the entire frequency band in terms of a series of power laws.  Applying this procedure to our  benchmark signals provides the reconstructions shown in Fig.~\ref{fig:recons}.  Notice that the intervals within which we perform the power-law reconstruction are not equally spaced. This is due to several reasons: among them, the SNR "density" (i.e.~the amount of SNR within a fixed frequency spacing) is not constant due to the frequency dependence of $\Omega_{\rm GW}$, and we must ensure that each bin contains a fraction of signal with sufficient SNR for being correctly reconstructed.


As Fig.~\ref{fig:recons} highlights, the  benchmark signals we consider can be reconstructed quite accurately. The uncertainty on the position of the peak and its amplitude is rather small. The detailed information that can be extracted by means of this multi-power-law reconstruction is matter of future investigations~\cite{CosWG2}. It is however clear that, if LISA detects a strong cosmological SGWB,  we will likely be able to decipher the origin of the signal. In particular, in case we detect a strong FOPT SGWB, from the reconstruction of the position and amplitude of the peak we can likely understand the energy scale at which the FOPT occurred, an outstanding information that can guide the particle physics community searching for the particle completion of the SM.
  
%
\section{Conclusions}
\label{sec:concl}

Several models of particle physics predict a cosmological FOPT. We  investigated the future prospect for the detection of the SGWB that such transitions produce. The results we  obtained apply to scenarios where the plasma plays a minor role in the production of the total SGWB: on the other hand,
our qualitative conclusions depend only marginally on this assumption.

We  considered the current PPTA, EPTA, NANOGRAV and aLIGO bounds on the SGWB, and shown that they poorly constrain the FOPT parameter space. On the contrary,  the parameter reach of the future GW network of SKA, LISA, ET and aLIGO-aVirgo-KAGRA is definitively more promising: by the end of the LISA mission,  this network will probe a large fraction of the FOPT parameter space. 
 If bubbles expand ultrarelativistically, any cosmological FOPT with $\beta/H_\star\lesssim 10^2$ and $10^{-3}\,\textrm{GeV}\lesssim T_\star \lesssim 10^{8}$\,GeV will be detected with SNR>10. The parameter space that can be probed remains considerable, although somehow reduced,  for subsonic bubbles.
In particular, due to the broadband of the FOPT SGWB, more than one detector is often sensitive to the same signal. There is then a manifest  synergy between  future GW experiments that will allow to probe the shape of a  SGWB signal at different frequencies, in case of a detection.The SGWB frequency shape carries valuable information about the processes that produced the signal~\footnote{Correlations between cosmological sources and signal shapes have been also studied in Refs.~\cite{Croon:2018erz, Kuroyanagi:2018csn}.}. To highlight this issue, we have considered some illustrative FOPT SGWBs with peaks inside the LISA  band. By means of a ``multi-bin" reconstruction procedure, we have shown that LISA has the potential to accurately reconstruct the signal. Recovering the position and amplitude of the signal peak  has rich implications. For instance (assuming a robust theoretical modellization of the FOPT signal) it allows to infer from the LISA data the FOPT contribution to the SGWB measured in other detectors (e.g.~SKA or ET) --  conclusion that those experiments could not reach independently if the signal is  dominated by other SGWB sources. The position and amplitude of the peak, moreover, allow one to estimate at which energy scale new physics beyond the  SM should emerge. This will be a key information for the collider community if no deviations from the SM will be found at the LHC.

%
\vspace{2mm}
{\em Acknowledgments:}
We thank the LISA CosWG for useful discussions. This work has been supported by the Ram\'on y Cajal Program of the Spanish MINEICO and the Universidad del Pa\'{\i}s Vasco UPV/EHU as a Visiting Professor (EM), the Spanish MINEICO under Grants FPA2015-64041-C2-1-P and FIS2017-85053-C2-1-P (EM), CICYT-FEDER-FPA2014-55613-P and FPA2017-88915-P (MQ), the Swiss National Science Foundation grant 200020-168988 (GN), the COST Action CA16104 ``Gravitational waves, black holes and fundamental physics" (GN), the Spanish MINEICOs ``Centro de Excelencia Severo Ocho'' Program grants SEV-2012-0249 (MP), and SEV-2016-0588 (MQ), the European Unions Horizon 2020 research and innovation programme under the Marie Sk\l{}odowska-Curie grant agreement No 713366 (MP), the STFC grant ST/P00055X/1 (GT), the Junta de Andaluc\'{\i}a under Grant FQM-225, and the Basque Government under Grant IT979-16 (EM).


\begin{thebibliography}{99}

\bibitem{Caprini:2015zlo}
  C.~Caprini {\it et al.},
  JCAP {\bf 1604} (2016) no.04,  001
  [arXiv:1512.06239 [astro-ph.CO]].

\bibitem{Bartolo:2016ami}
  N.~Bartolo {\it et al.},
  JCAP {\bf 1612} (2016) no.12,  026
  [arXiv:1610.06481 [astro-ph.CO]].
  

\bibitem{Caprini:2018mtu}
  C.~Caprini and D.~G.~Figueroa,
  arXiv:1801.04268 [astro-ph.CO].
  
  
\bibitem{Kajantie:1996mn}
  K.~Kajantie, M.~Laine, K.~Rummukainen and M.~E.~Shaposhnikov,
  Phys.\ Rev.\ Lett.\  {\bf 77} (1996) 2887
  [hep-ph/9605288].
  
  
  
  
\bibitem{Huber:2008hg}
  S.~J.~Huber and T.~Konstandin,
  JCAP {\bf 0809} (2008) 022
  [arXiv:0806.1828 [hep-ph]].

  
  
  
\bibitem{Cutting:2018tjt}
  D.~Cutting, M.~Hindmarsh and D.~J.~Weir,
  arXiv:1802.05712 [astro-ph.CO].

 
  
\bibitem{Shannon:2015ect}
  R.~M.~Shannon {\it et al.},
  Science {\bf 349} (2015) no.6255,  1522
  [arXiv:1509.07320 [astro-ph.CO]].
  
  
  
  
\bibitem{Lentati:2015qwp}
  L.~Lentati {\it et al.},
  Mon.\ Not.\ Roy.\ Astron.\ Soc.\  {\bf 453} (2015) no.3,  2576
  [arXiv:1504.03692 [astro-ph.CO]].
  
  
  
\bibitem{TheLIGOScientific:2016dpb}
  B.~P.~Abbott {\it et al.} [LIGO Scientific and Virgo Collaborations],
  Phys.\ Rev.\ Lett.\  {\bf 118} (2017) no.12,  121101
   Erratum: [Phys.\ Rev.\ Lett.\  {\bf 119} (2017) no.2,  029901]
  [arXiv:1612.02029 [gr-qc]].
 

\bibitem{Arzoumanian:2018saf}
  Z.~Arzoumanian {\it et al.} [NANOGRAV Collaboration],
  arXiv:1801.02617 [astro-ph.HE].

  
\bibitem{Audley:2017drz}
  H.~Audley {\it et al.} [LISA Collaboration],
  arXiv:1702.00786 [astro-ph.IM].


\bibitem{Aasi:2013wya}
  B.~P.~Abbott {\it et al.} [KAGRA and LIGO Scientific and VIRGO Collaborations],
  Living Rev.\ Rel.\  {\bf 21} (2018) 3
   [Living Rev.\ Rel.\  {\bf 19} (2016) 1]
  [arXiv:1304.0670 [gr-qc]].

 
 
\bibitem{Sathyaprakash:2012jk}
  B.~Sathyaprakash {\it et al.},
  Class.\ Quant.\ Grav.\  {\bf 29} (2012) 124013
   Erratum: [Class.\ Quant.\ Grav.\  {\bf 30} (2013) 079501]
  [arXiv:1206.0331 [gr-qc]].
  
\bibitem{Moore:2014lga}
  C.~J.~Moore, R.~H.~Cole and C.~P.~L.~Berry,
  Class.\ Quant.\ Grav.\  {\bf 32} (2015) no.1,  015014
  [arXiv:1408.0740 [gr-qc]].
 

 
\bibitem{Megias:2018sxv}
  E.~Megias, G.~Nardini and M.~Quiros,
  JHEP {\bf 1809} (2018) 095
  [arXiv:1806.04877 [hep-ph]].
  
 

\bibitem{Evans:2016mbw}
  B.~P.~Abbott {\it et al.} [LIGO Scientific Collaboration],
  Class.\ Quant.\ Grav.\  {\bf 34} (2017) no.4,  044001
  [arXiv:1607.08697 [astro-ph.IM]].
 
 
 
\bibitem{Axen:2018zvb}
  M.~F.~Axen, S.~Banagiri, A.~Matas, C.~Caprini and V.~Mandic,
  arXiv:1806.02500 [astro-ph.IM].
 
 
\bibitem{CosWG1}
   LISA Cosmology Working Group, to appear.
   

\bibitem{Adams:2013qma}
  M.~R.~Adams and N.~J.~Cornish,
  Phys.\ Rev.\ D {\bf 89} (2014) no.2,  022001
  [arXiv:1307.4116 [gr-qc]].
  

\bibitem{CosWG2}
   LISA Cosmology Working Group, in progress.
     
     
\bibitem{Croon:2018erz}
  D.~Croon, V.~Sanz and G.~White,
  arXiv:1806.02332 [hep-ph].
  
\bibitem{Kuroyanagi:2018csn}
  S.~Kuroyanagi, T.~Chiba and T.~Takahashi,
  arXiv:1807.00786 [astro-ph.CO].     

\end{thebibliography}
\end{document}